\documentclass[conference]{IEEEtran}
\IEEEoverridecommandlockouts
\usepackage{cite}
\usepackage{amsmath,amssymb,amsfonts}
\usepackage{algorithmic}
\usepackage{graphicx}
\usepackage{textcomp}
\usepackage{xcolor}
\usepackage{booktabs}
\usepackage{comment}

\usepackage{listings}
\lstdefinestyle{lf}{
    language=C,
    basicstyle=\ttfamily\footnotesize,
    keywordstyle=\color{blue}\bfseries,
    commentstyle=\color{green!60!black}\itshape,
    stringstyle=\color{red},
    numbers=none,
    numberstyle=\tiny\color{gray},
    backgroundcolor=\color{gray!5},
    frame=single,
    breaklines=true,
    captionpos=b,
    xleftmargin=2pt,
    xrightmargin=2pt
}

\def\BibTeX{{\rm B\kern-.05em{\sc i\kern-.025em b}\kern-.08em
    T\kern-.1667em\lower.7ex\hbox{E}\kern-.125emX}}
\begin{document}

\title{Ensuring Deterministic Timing in a Federated GNSS Correction Pipeline with Lingua Franca}

\author{
\IEEEauthorblockN{Tejeswini Jayaramareddy}
\IEEEauthorblockA{
\textit{Santa Clara University}\\
Santa Clara, CA, United States\\
tjayaramareddy@scu.edu}
\and
\IEEEauthorblockN{Hokeun Kim}
\IEEEauthorblockA{
\textit{Arizona State University}\\
Tempe, AZ, United States\\
hokeun@asu.edu}
\and
\IEEEauthorblockN{Hoeseok Yang}
\IEEEauthorblockA{
\textit{Santa Clara University}\\
Santa Clara, CA, United States\\
hoeseok.yang@scu.edu }
}

\maketitle

\begin{abstract}
Embedded systems that combine hardware interrupts, buffering, and distributed communication are often perceived as inherently asynchronous and difficult to analyze. However, such systems can exhibit a deterministic timing structure when modeled using explicit logical-time semantics.

This paper presents a Global Navigation Satellite System (GNSS) correction-data pipeline implemented as a federated Lingua Franca (LF) application. The federated LF program decomposes the end-to-end pipeline into reactors with explicit time semantics, including a time-triggered GNSS receiver, a UART interrupt stream derived from baud rate and First-In First-Out (FIFO) buffer characteristics, a periodic forwarding task, and downstream processing with jitter monitoring.

Federated execution and runtime logs validate the analytically derived deterministic timing structure—including interrupt cadence, ring-buffer evolution, packetization behavior, and physical--logical jitter—yielding a reproducible and predictable timing profile.
\end{abstract}


\section{Introduction}

Global Navigation Satellite System (GNSS) correction-data pipelines are core components of modern automotive systems~\cite{shi2024application}. In typical deployments, a GNSS receiver generates binary correction messages, such as Radio Technical Commission for Maritime Services (RTCM)~\cite{rtcm_ndf}, at deterministic epoch intervals with payload sizes varying according to constellation configuration and satellite visibility. This data stream is transported via a Universal Asynchronous Receiver/Transmitter (UART) interface to a microcontroller (MCU)~\cite{infineon_tc3xx_manual_part1} and forwarded over automotive Ethernet to high-performance processors executing Position Engine algorithms~\cite{rodriguez2021protection}. The pipeline must ensure deterministic buffering, bounded latency, and loss-free transmission under real-time constraints, as timing jitter or data loss degrades positioning accuracy and downstream vehicle control performance.

Although each hardware interface is configured deterministically (e.g., baud rate, interrupt thresholds, periodic timers), end-to-end behavior appears asynchronous due to interrupt-driven reception, First-In First-Out (FIFO) buffering, DMA transfers, and distributed execution across heterogeneous processing units. This complicates reasoning about latency bounds and causality, motivating a modeling abstraction that makes timing and execution ordering explicit rather than implicit in operating system scheduling.

To address this challenge, we adopt the \emph{reactor model}, which structures systems as composable components reacting to events and timers under explicit time semantics. Instead of threads and preemptive scheduling, reactors define concurrency through causality and timestamped events. Reactions occur at discrete logical times, and execution order follows data dependencies rather than runtime interleavings, enabling deterministic composition and analyzability.

Building on this foundation, we employ \emph{Lingua Franca (LF)}~\cite{lohstroh2021toward,lfdocs}, a coordination language providing deterministic logical-time semantics. In LF, timers, physical actions (e.g., UART interrupts), and network delays are declared directly in the program. Logical time is separated from physical execution time, ensuring uniquely defined behavior for a given sequence of timestamped inputs. LF supports federated execution, allowing distributed components to run as separate processes while preserving causality and logical-time consistency across network boundaries. This paper makes determinism explicit in an interrupt-driven, federated GNSS pipeline and shows how its timing structure can be formally derived for predictable design~\cite{lin2023towards}.

This paper makes the following contributions:

\begin{itemize}
\item We present a federated LF model of a GNSS correction pipeline, capturing UART interrupts, buffering, and distributed forwarding under logical-time semantics.

\item We derive closed-form characterizations of deterministic timing as functions of system parameters (baud rate, payload size, FIFO depth, epoch period).

\item We experimentally validate the derived timing structure across 317 epochs, confirming bounded latency and stable buffer behavior.

\end{itemize}

\section{GNSS Correction-Data Pipeline}
\label{sec:system_arch}

\begin{figure*}[t]
\centering
\includegraphics[width=\textwidth, height=4.8cm]{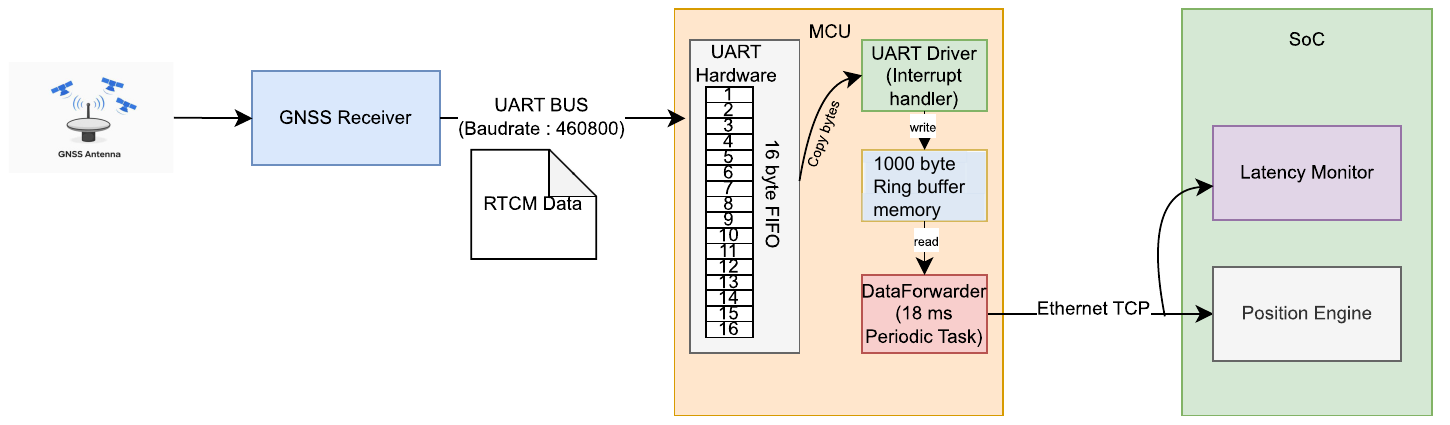}
\caption{GNSS correction-data pipeline architecture}
\label{fig:SystemArchitecture}
\end{figure*}

Fig.~\ref{fig:SystemArchitecture} depicts the GNSS correction-data pipeline considered in this work at the architectural level.

In contrast to the high-level description provided in the Introduction, this section formalizes the timing behavior of each stage in the pipeline. The system consists of a time-triggered GNSS receiver, an interrupt-driven UART interface with FIFO buffering, and a periodic forwarding task executing on the MCU before data is processed on the SoC.

While these components operate under individually deterministic parameters (e.g., fixed epoch period, baud rate, and task periods), their interaction leads to a non-trivial timing structure. In particular, the combination of interrupt-driven data ingestion and periodic forwarding introduces phase-dependent behavior that is not immediately apparent from the implementation.

In the following subsections, we model each stage of the pipeline and derive its timing characteristics, forming the basis for the deterministic analysis presented in this paper.



\subsection{GNSS Receiver}

The GNSS receiver firmware operates with a fixed epoch period of $T_{gnss} = 100~\text{ms}$.
At each epoch, RTCM3~\cite{rtcm2016rtcm} correction messages are generated with bounded payload size.
Due to variations in satellite visibility and constellation configuration, 
realistic MSM (Multiple Signal Messages) sizes range between 400 and 800 bytes, i.e., $D \in [400, 800]$~bytes.
Upon generation, the payload is transmitted immediately over a UART interface 
configured at $460,800$ baud.

\subsection{UART Interface and MCU Ingestion}

Assuming 10 bits per byte (1 start bit, 8 data bits, and 1 stop bit), the byte transmission time is
\[
t_{byte} = \frac{10}{460800} \approx 21.7~\mu s.
\]

In the worst case, where $D = 800$ bytes, the maximum UART transfer time becomes 
\[T_{uart}^{max} = 800 \cdot 21.7~\mu s \approx 17.4~\text{ms}
\]

The UART peripheral includes a 16-byte hardware FIFO. An interrupt is generated when the FIFO reaches the configured threshold, producing interrupt cadence of
\[
T_{irq} = 16 \cdot t_{byte} \approx 347~\mu s.
\]

Upon each interrupt, the interrupt service routine (ISR) copies 16-byte chunks from the FIFO into a 1000-byte ring buffer implemented with separate write and read pointers.

\subsection{Periodic Forwarding and SoC Processing}

The MCU executes a periodic \textit{DataForwarder} task with period $T_{read} = 18~\text{ms}$.
At each activation, all available bytes are drained from the ring buffer
and transmitted over Ethernet (TCP) to the SoC, where the PE~\cite{rodriguez2021protection} processes the incoming RTCM stream.

Because $T_{gnss}=100$~ms and $T_{read}=18$~ms are non-harmonic,
their relative phase evolves deterministically over time.
The joint schedule repeats every super-cycle period
\[
T_{super} = \mathrm{LCM}(100,18) = 900~\text{ms}.
\]

Forwarding events occur at $t_k = k\cdot T_{read}$, while GNSS epochs occur at
$e_n = n\cdot T_{gnss}$.
The inter-arrival timing pattern at the SoC is therefore governed by the modular phase
\[
\phi_k = (k\cdot T_{read}) \bmod T_{gnss}.
\]

Since $\gcd(100,18)=2$, the phase sequence repeats every
$900$ ms. A payload generated at epoch $e_n$ becomes observable at the
first forwarding instant satisfying
\[
t_k \ge e_n + T^{\max}_{uart}.
\]
With $T^{\max}_{uart}=17.4$~ms, evaluation over one super-cycle
yields the closed inter-arrival set
\[
\Delta \in \{18,\,90,\,108\}\text{ ms}.
\]
Thus, the inter-arrival timing is determined by phase alignment between $T_{gnss}=100$\,ms and $T_{read}=18$\,ms. 
For example: Epoch~1 ($t=0$) forwards at 18\,ms ($\Delta=18$\,ms); 
Epoch~2 ($t=100$) at 108\,ms ($\Delta=90$\,ms); 
Epoch~3 ($t=200$) at 216\,ms ($\Delta=108$\,ms), repeating over the 900\,ms super-cycle.

Worst-case latency occurs when payload completion happens immediately after a forwarding activation, requiring the data to wait an additional forwarding period before transmission. The resulting upper bound is
\[
L_{\max} = T_{uart}^{\max} + T_{read}
= 17.4~\text{ms} + 18~\text{ms}
= 35.4~\text{ms}.
\]

Table~\ref{tab:timeline} summarizes the deterministic event sequence within first 100\,ms epoch. 


\begin{table}[t]
\caption{System Timeline for First Epoch ($t = 0$--100 ms)}
\label{tab:timeline}
\centering
\renewcommand{\arraystretch}{1.05}
\setlength{\tabcolsep}{4.5pt}
\begin{tabular}{r|c|l}
\hline
\textbf{Time ($\mu$s)} & \textbf{Event/Activity} & \textbf{Component} \\ 
\hline
0 & GNSS epoch triggers (400--800 B) & GNSSReceiver \\ 
\hline
0 & UART reception begins & UARTDriver \\ 
\hline
347 & FIFO interrupt \#1 (16 B copied) & UARTDriver \\ 
\hline
694 & FIFO interrupt \#2 (16 B copied) & UARTDriver \\ 
\hline
1041 & FIFO interrupt \#3 (16 B copied) & UARTDriver \\ 
\hline
$\cdots$ & Additional FIFO interrupts (25--50/epoch) & UARTDriver \\ 
\hline
8700--17400 & UART transfer complete & UARTDriver \\ 
\hline
18000 & Periodic drain triggered & DataForwarder \\ 
\hline
18000 & 400--800 B transmitted & DataForwarder \\ 
\hline
18000 & RTCM received & PositionEngine \\ 
\hline
36000--90000 & Periodic drains (no data) & DataForwarder \\ 
\hline
100000 & Next GNSS epoch & GNSSReceiver \\ 
\hline
\end{tabular}
\end{table}

\section{Lingua Franca Model and Implementation}

\begin{figure}[t]
\centering
\includegraphics[width=0.85\columnwidth,height=0.4\columnwidth,keepaspectratio]{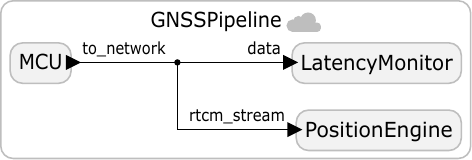}
\caption{GNSS correction-data pipeline architecture implemented using LF.}
\label{fig:GNSSPipeline}
\end{figure}

This section describes how the GNSS correction pipeline introduced in the previous section is specified as a federated LF program. As illustrated in Fig.~\ref{fig:GNSSPipeline}, the \texttt{GNSSPipeline} reactor formalizes the end-to-end data path using federated reactors under LF’s deterministic logical-time model. 

\begin{figure}
\begin{lstlisting}[style=lf,caption={Top-Level Federation}]
federated reactor GNSSPipeline {
    mcu = new MCU()
    soc = new PositionEngine()
    monitor = new LatencyMonitor()
    mcu.to_network -> soc.rtcm_stream
    mcu.to_network -> monitor.data
}
\end{lstlisting}
\end{figure}

The top-level \texttt{GNSSPipeline} federated reactor instantiates these federates and declares their communication channels. The \textit{MCU} federate produces correction data, which is transmitted to both the \textit{PositionEngine} for processing and the \textit{LatencyMonitor} for runtime measurement and verification.
Each federate executes as a separate process under LF’s
decentralized runtime. Logical time is synchronized across
network boundaries with bounded delay.

\begin{figure}[t]
\centering
\includegraphics[width=9.2cm, height=3cm]{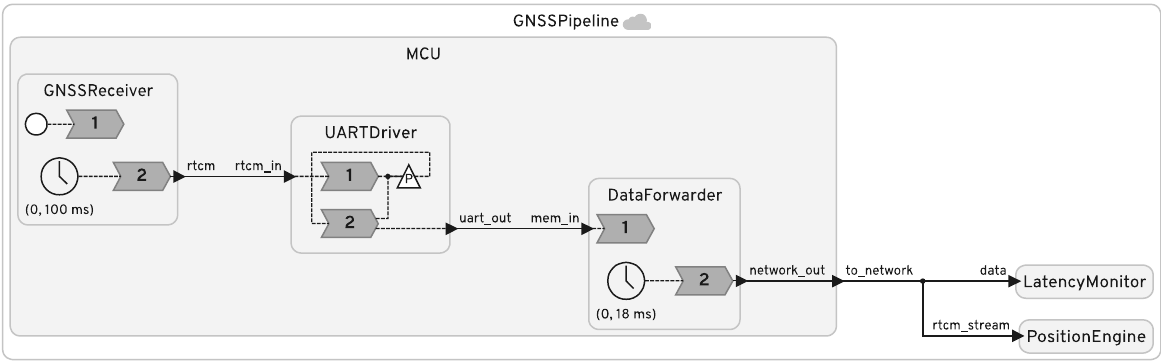}
\caption{Detailed MCU implementation showing UART driver, hardware FIFO, buffering, and task interactions.}
\label{fig:MCU}
\end{figure}

\noindent\textbf{MCU Composite Reactor}:
As illustrated in Fig.~\ref{fig:MCU}, the MCU federate is modeled as a composite reactor that makes the internal data path explicit. It encapsulates three subordinate reactors—\textit{GNSSReceiver}, \textit{UARTDriver}, and \textit{DataForwarder}—which together represent epoch generation, interrupt-driven ingestion, and periodic forwarding, respectively.

\begin{lstlisting}[style=lf,caption={MCU Composition}]
reactor MCU {
    output to_network: int
    gnss = new GNSSReceiver()
    uart = new UARTDriver()
    forwarder = new DataForwarder()
    gnss.rtcm -> uart.data_in
    uart.uart_out -> forwarder.mem_in
    forwarder.network_out -> to_network
}
\end{lstlisting}

The internal connections explicitly encode the data-flow dependencies: RTCM payloads generated by the GNSS receiver are delivered to the UART driver, buffered and propagated to the forwarding task, and ultimately emitted to the network interface. Consequently, execution ordering is determined by logical causality and data dependencies rather than operating-system scheduling or thread interleavings.

\noindent\textbf{GNSS Receiver Reactor}:
The \texttt{GNSSReceiver} reactor models the time-triggered generation of RTCM correction messages at a fixed epoch period as shown below.

\begin{lstlisting}[style=lf, caption=GNSS Receiver Implementation]
reactor GNSSReceiver {
  output rtcm: int
  timer epoch(0, 100 msec)
  state message_count: int = 0
  state min_bytes: int = 400
  state max_bytes: int = 800
  reaction(startup) {=
    srand(time(NULL));  =}
  reaction(epoch) -> rtcm {=
    self->message_count++;
    // Variable size models real-world conditions
    int bytes_this_epoch = self->min_bytes + 
        (rand() % (self->max_bytes - self->min_bytes + 1));

    lf_set(rtcm, bytes_this_epoch);
  =}
}
\end{lstlisting}


The declaration \texttt{timer epoch(0, 100 msec)} specifies a strictly periodic logical timer that fires at $t = 0, 100, 200, \ldots$ ms without drift. These firings are defined in logical time and are therefore independent of physical execution latency or scheduling delays.
The payload size is sampled uniformly within the bounded interval $[400, 800]$ bytes to emulate variability due to satellite visibility and constellation conditions. Importantly, while the data value varies, the logical timestamp associated with each epoch remains deterministic. As a result, variability is confined to payload magnitude, whereas temporal behavior is governed entirely by explicit logical-time semantics. The physical action marks the boundary between unpredictable hardware time and deterministic logical
execution inside the reactor network.

\noindent\textbf{UART Driver}:
Interrupt-driven UART behavior is modeled using a
\textit{physical action}, which explicitly separates hardware
timing from logical-time semantics.

The physical action marks the boundary between unpredictable hardware time and deterministic logical execution inside the reactor network.

\begin{figure}[t]
\begin{samepage}
\begin{lstlisting}[style=lf,caption={UART Driver with Physical Action}]
reactor UARTDriver {
    input data_in: int
    physical action hw_interrupt: int
    state buffer_used: int = 0
    reaction(data_in) -> hw_interrupt {=
        int interrupts = (data_in->value + 15) / 16;
        for (int i = 0; i < interrupts; i++) {
            lf_schedule_int(hw_interrupt,
                            USEC(347), 16);
        }   =}
    reaction(hw_interrupt) {=
        self->buffer_used += hw_interrupt->value;=}
}
\end{lstlisting}
\end{samepage}
\end{figure}

\noindent\textbf{DataForwarder}:
The \texttt{DataForwarder} reactor models the periodic transmission on the MCU, decoupling interrupt-driven buffer writes from time-triggered network forwarding by introducing an explicit logical timer. Because $T_{gnss}\!=\!100$\,ms and $T_{read}\!=\!18$\,ms are non-harmonic, the periodic reactor induces deterministic phase evolution across federates.

\begin{figure}[t]
\begin{lstlisting}[style=lf,caption={DataForwarder Reactor}]
reactor DataForwarder {
    input mem_in: int
    output network_out: int
    timer period(0, 18 msec)
    reaction(mem_in) {=
        /* interrupt-driven writes */ =}
reaction(period) -> network_out {=
        /* periodic drain of buffer  */ =}
}
\end{lstlisting}
\end{figure}

\textbf{Logical-Physical Alignment.} While defined in logical time,
the LF runtime executes reactions at $t+\epsilon$, where $\epsilon$ is bounded.
Measured deviations are $<\!10\,\mu$s (mean 4.3~$\mu$s), i.e.,
$<\!0.01\%$ of the 100\,ms period. This negligible jitter does not
affect buffer or forwarding decisions; thus, logical properties carry over to physical execution.
\section{Experimental Results}
\label{sec:results}

\noindent\textbf{Experimental Setup}:
The federated LF implementation was executed on Linux
(Ubuntu 22.04, Intel i7-9700K, 16\,GB RAM) using the
parameters defined in Section~\ref{sec:system_arch}.
Over 31.7\,s, 317 GNSS epochs were processed with
payloads sampled uniformly in $D\in[400,800]$ bytes.

\noindent\textbf{Ring Buffer Occupancy}:
The ring buffer exhibits the expected burst-and-drain behavior induced by interrupt-driven FIFO fills and periodic forwarding. Peak occupancy remained below the analytical bound ($B = 1000$ bytes), and no overflow events were observed. Table II summarizes the detailed buffer statistics, including peak utilization and mean occupancy, confirming bounded and stable memory usage.

\begin{table}[t]
\centering
\caption{Ring Buffer Statistics}
\label{tab:buffer}
\renewcommand{\arraystretch}{1.05}
\setlength{\tabcolsep}{5pt}
\begin{tabular}{lrl}
\toprule
\textbf{Metric} & \textbf{Value} & \textbf{Unit} \\
\midrule
Capacity ($B$) & 1{,}000 & bytes \\
Peak occupancy & 816 & bytes \\
Peak utilization & 81.6 & \% \\
Mean occupancy & 287.3 & bytes \\
Overflow events & 0 & count \\
\bottomrule
\end{tabular}
\end{table}

\noindent\textbf{Inter-Arrival Timing}:
Arrivals form the closed set $\{18,90,108\}$ ms, as analytically
derived in Section~II. Across 317 epochs, the distribution was
19\%, 41\%, and 40\%, with identical logical timestamps across runs.


\noindent\textbf{End-to-End Latency}:
Measured latency from GNSS epoch to PositionEngine reception exhibits a multimodal distribution reflecting the deterministic $T_{gnss}$–$T_{read}$ phase interaction (Section~\ref{sec:system_arch}).
The maximum closely matches the 35.4\,ms theoretical bound. The slight excess arises from physical runtime overhead outside the logical-time model, while logical timestamps remain deterministic. This delay can be treated as a bounded, platform-dependent margin to the analytical bound in practice. Detailed statistics are summarized in Table III.

\begin{table}[t]
\centering
\caption{End-to-End Latency Statistics}
\label{tab:latency}
\renewcommand{\arraystretch}{1.05}
\setlength{\tabcolsep}{5pt}
\begin{tabular}{lrl}
\toprule
\textbf{Metric} & \textbf{Value} & \textbf{Unit} \\
\midrule
Mean & 22.4 & ms \\
P95/P99 & 35.2/36.8 & ms \\
Max (measured) & 37.1 & ms \\
Bound (theory) & 35.4 & ms \\
\bottomrule
\end{tabular}
\end{table}

\section{Conclusion}

This paper presented a federated Lingua Franca (LF) model of an interrupt-driven GNSS correction pipeline with explicit timing parameters ($T_{gnss}=100$\,ms, $T_{read}=18$\,ms, 460{,}800 baud, $D \in [400,800]$ bytes). Although such pipelines appear asynchronous due to interrupts, buffering, and distributed execution, we showed that their timing behavior is analytically derivable under logical-time semantics.

From these parameters, we derived the 900\,ms super-cycle, the closed inter-arrival set $\{18, 90, 108\}$\,ms, and the worst-case latency bound $L_{\max}=35.4$\,ms. Experiments over 317 epochs confirmed identical logical traces, bounded latency, stable buffer occupancy, and zero overflows, closely matching theoretical predictions. 

These results demonstrate that logical-time modeling renders an interrupt-driven pipeline predictable and analytically tractable, enabling reproducible and formally analyzable real-time behavior.

\bibliographystyle{IEEEtran} 
\bibliography{ref} 

\end{document}